\documentclass[a4paper,12pt]{article} 
 \pdfoutput=1 
\usepackage{amsmath}
\usepackage{bm,color,url}
\usepackage{amssymb}
\usepackage{graphicx}

\begin{document}
\titlepage

\today

  \begin{flushright}
   {\bf
  \begin{tabular}{l}
OCHA-PP-344
  \end{tabular}
   }
  \end{flushright}

\vspace*{1.0cm}
\baselineskip 18pt
\begin{center}
{\Large \bf 
Search for lepton flavor violation at future lepton colliders
} 

\vspace*{1.0cm} 

{\bf 
 Gi-Chol Cho$^a$, Hanako Shimo$^b$
}

\vspace*{0.5cm}
$^a${\em Department of Physics, Ochanomizu University, Tokyo 112-8610, Japan} \\
$^b${\em Graduate school of Humanities and Sciences, Ochanomizu University, Tokyo 112-8610, Japan} \\
\end{center}

\vspace*{1cm}

\baselineskip 18pt
\begin{abstract}
 \noindent
Lepton flavor violating (LFV) processes, $e^+ e^- \to e^+ \ell^-$ and $e^- e^-
 \to e^- \ell^-$ ($\ell=\mu$ or $\tau$), via four-Fermi contact 
 interactions at future International Linear Collider (ILC) are studied.
  Taking account of previous experimental results of LFV processes
 $\mu\to 3e$ and 
 $\tau \to 3e$, we find that the upper limits on the LFV parameters for
 $\ell=\tau$ could be improved at the ILC experiment using the 
 polarized electron beam. The improvement of the upper limits could be 
 nearly an order of magnitude as compared to previous ones. 
\end{abstract}

\newpage
\section{Introduction} 
Discovery of the neutrino oscillation~\cite{Fukuda:1998mi} implies the
finite but tiny mass of neutrinos and opens a window to new physics
beyond the Standard Model.
One of the direct consequences of massive neutrinos 
is the lepton-flavor violating (LFV) 
processes, though the size of the signal is highly model dependent.
For example, a simple extension of the SM which
allows neutrinos to be massive 
predicts the branching ratio of $\mu \to e\gamma$ as 
$
\mathrm{Br}(\mu\to e \gamma)
=\frac{3}{32\pi}\alpha |U^*_{ei}U_{\mu i}|^2
(m_{\nu_i}/m_W)^4< 10^{-48}(m_{\nu_i}/1~\mathrm{eV})^4
$, which is hopeless to be observed
(where $\alpha, U_{ij}$ and $i$ denote the fine-structure constant,
 lepton-flavor mixing matrix and the generation index, respectively).
 In some class of new physics models, however, there could be sources of
sizable LFV.
Origin of the LFV and phenomenological predictions 
have been extensively studied based 
on various ideas such as supersymmetry~\cite{Barbieri:1994pv,
Barbieri:1995tw, Hisano:1995nq, Hisano:1995cp},
extra-dimension~\cite{Chang:2005ag, Agashe:2006iy,Csaki:2008qq}, 
and so on~\cite{deGouvea:2013zba}. 

Although no evidence of the LFV has been observed so far, 
several experiments are aiming to find signatures of LFV in the
charged lepton sector, e.g.,
$\mu \to e\gamma$ at MEG II~\cite{Baldini:2013ke},
$\mu \to e e e$ at Mu3e~\cite{Bravar:2015vja}, 
$\mu$-$e$ conversion ($\mu N \to e N$) at 
COMET~\cite{Kuno:2013mha} and Mu2e~\cite{Bartoszek:2014mya}, 
the LFV in $\tau$ decays at Belle II~\cite{Aushev:2010bq}, etc
(see e.g. \cite{Marciano:2008zz, Bernstein:2013hba} for reviews). 
The sensitivities of (some of) these experiments to new physics search
have been studied in, e.g., refs.~\cite{Koike:2010xr, Uesaka:2016vfy,
Crivellin:2013hpa}, based on the effective Lagrangian with higher
dimensional operators.
In this paper, we investigate a possibility to search for the LFV
processes from the four-Fermi contact interactions at the collider
experiments.
We focus the processes 
\begin{align}
 e^+ e^- &\to e^+ \ell^-,
 \label{signalepem}
 \\
 e^- e^- &\to e^- \ell^-,
 \label{signalemem}
\end{align}
for $\ell=\mu,\tau$ 
at future $e^+ e^-$ linear collider (International Linear Collider,
ILC)~\cite{Baer:2013cma} and $e^- e^-$ collider as its option\footnote{
A possibility of the LFV process (\ref{signalepem}) due to the sterile
neutrino production and decay has been discussed in 
ref.~\cite{Antusch:2016ejd}, while 
the process (\ref{signalemem}) has been studied in the type-II seesaw 
model in ref.~\cite{Rodejohann:2010bv}.
}. 
The new physics effects on these processes can be parametrized by six
couplings for each lepton flavor $\ell$, and we examine constraints on
these
parameters from the ILC experiments. 
We show that the ILC experiment is less sensitive than the 
previous LFV experiments for $\ell=\mu$ case.
On the other hand, 
upper bounds on the LFV parameters could be improved by more than an 
order of magnitude from the previous bound for $\ell=\tau$ case,
in particular by using the polarized electron beam.
This paper is organized as follows. We briefly review the effective
Lagrangian of four-Fermi contact interactions and some observables
related to our study in Sec.~II. Numerical analysis and limits on the
LFV parameters will be given in Sec.~III. Sec.IV is devoted to
summary. 

\section{Four-fermi interactions and observables}

The effective interaction Lagrangian describes the LFV processes via
the four-Fermi contact interactions is given by~\cite{Kuno:1999jp} 
\begin{align}
 {\mathcal{L}_{\mathrm{eff}}}&=
 -\frac{4G_F}{\sqrt{2}}
 \left\{
  g^\ell_1 \left(\overline{\ell_R}e_L \right) \left(\overline{e_R}e_L\right)
   +
  g^\ell_2 \left(\overline{\ell_L}e_R \right) \left(\overline{e_L}e_R \right)		  
  \right.
\nonumber \\
 &+
  g^\ell_3 \left(\overline{\ell_R}\gamma^\mu e_R \right)
  \left(\overline{e_R}\gamma_\mu e_R \right)		  
  +
  g^\ell_4 \left(\overline{\ell_L}\gamma^\mu e_L \right)
  \left(\overline{e_L}\gamma_\mu e_L \right)		  
\nonumber \\
 &+
 \left.
  g^\ell_5 \left(\overline{\ell_R}\gamma^\mu e_R \right)
  \left(\overline{e_L}\gamma_\mu e_L \right)		  
  +
  g^\ell_6 \left(\overline{\ell_L}\gamma^\mu e_L \right)
  \left(\overline{e_R}\gamma_\mu e_R \right)		  
 \right\}+\mathrm{h.c.},
 \label{eq:efflag}
\end{align}
where the Fierz rearrangement is used.
The suffix $\ell$ stands for $\mu$ or $\tau$, 
and $G_F$ denotes the Fermi coupling constant.
The subscripts $L$ 
and $R$ represent the chirality of a fermion $f$, i.e., $f_{L(R)}\equiv 
\frac{1-(+)\gamma_5}{2}f$. 
There are six dimensionless couplings $g_i~(i=1\sim 6)$ but only three
parameters are constrained from the LFV experiments as will be shown
later.

In the limit of massless leptons, 
the spin-averaged differential cross-section in the center-of-mass (CM) 
system for $e^+ e^- \to e^+ \ell^-$ and $e^- e^- \to e^- \ell^-$ are
calculated from the effective Lagrangian~(\ref{eq:efflag}) as  
\begin{align}
 \frac{d \sigma(e^+ e^- \to e^+ \ell^-)}{d\cos\theta}
  &=
  \frac{G_F^2 s}{64\pi}
  \left[
   \left( G^\ell_{12}+16G^\ell_{34} \right) (1+\cos\theta)^2
  +4G^\ell_{56}\left\{4+(1-\cos\theta)^2 \right\}
	   \right], 
\label{diffxepem}
  \\
 \frac{d \sigma(e^- e^- \to e^- \ell^-)}{d\cos\theta}
  &=
  \frac{G_F^2 s}{16\pi}
  \left[
   G^\ell_{12}+16G^\ell_{34} +2G^\ell_{56}(1+\cos\theta)^2 
				\right],
\label{diffxemem}  
\end{align}
where $\sqrt{s}$ denotes the total energy in the CM-system and 
the parameter $G^\ell_{ij}$ is defined as
\begin{align}
 G^\ell_{ij}\equiv |g^\ell_i|^2+  |g^\ell_j|^2.  
\end{align}

\begin{figure}[t]
  \begin{center}
   \includegraphics[clip, width=16cm]{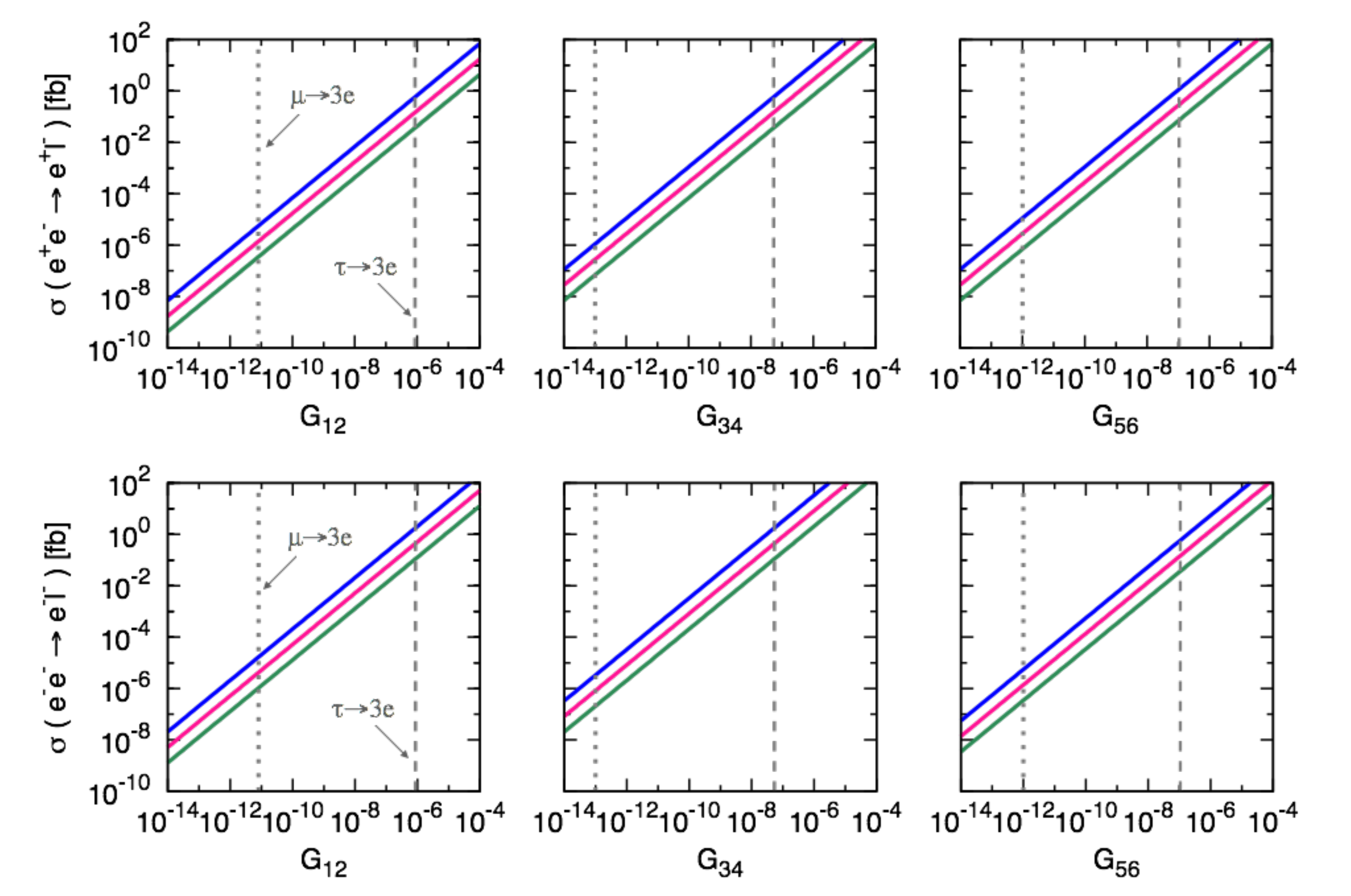}
   \caption{The cross section of $e^+ e^- \to e^+ \ell^-$ (upper) and
   $e^- e^- \to e^- \ell^-$ (lower) for $\sqrt{s}=250~\mathrm{GeV}$
   (green), $500~\mathrm{GeV}$ (red) and $1~\mathrm{TeV}$ (blue).
   The dotted and dashed vertical lines denote the upper bounds on
   $G_{ij}$ from $\mu \to 3e$~\cite{Bellgardt:1987du} and 
   $\tau \to 3e$~\cite{Hayasaka:2010np}.  
   }
    \label{fig:xsec_epem}
  \end{center}
\end{figure}
The couplings in the effective Lagrangian~(\ref{eq:efflag}) also induce 
the LFV process $\mu \to eee$ or $\tau \to eee$ 
(hereafter we denote these processes as $\mu\to 3e$ and $\tau \to 3e$
for simplicity). 
The branching ratio of $\mu \to 3e$ is expressed in terms of $G^\mu_{ij}$
as~\cite{Okada:1999zk}
\begin{align}
 \mathrm{Br}(\mu \to 3e)
&= \frac{\Gamma(\mu\to 3e)}{\Gamma(\mu \to e \nu_\mu \bar{\nu}_e)}
= \frac{1}{8}\left(G^\mu_{12}+16 G^\mu_{34}+ 8G^\mu_{56}\right),
\label{brmu3e}
\end{align}
while that of $\tau \to 3e$ is 
\begin{align}
\mathrm{Br}(\tau\to 3e)
&=\frac{\tau_\tau}{\tau_\mu}
\left(\frac{m_\tau}{m_\mu}\right)^5
\times \frac{1}{8}
\left(G^\tau_{12}+16 G^\tau_{34}+8 G^\tau_{56}\right)
\nonumber \\
 &\approx 0.022 \times \left(G^\tau_{12}+16 G^\tau_{34}+8 G^\tau_{56}\right),
  \label{brtau3l}
\end{align}
where $\tau_\tau$ and $\tau_\mu$ are the lifetime of $\tau$ and $\mu$,
respectively, and we adopt $\tau_\tau=2.91 \times 10^{-13}~\mathrm{s}$
and $\tau_\mu=2.20 \times 10^{-6}~\mathrm{s}$ for the numerical
evaluation~\cite{Agashe:2014kda}.
To find constraints on the LFV parameter $G_{ij}$ from $\mu\to 3e$ and
$\tau \to 3e$, we summarize current experimental bounds on those
processes.
The upper bounds on $\mathrm{Br}(\mu^+ \to e^+ e^+ e^-)$ and
$\mathrm{Br}(\tau \to 3\ell)$
have been given by the SINDRUM~\cite{Bellgardt:1987du}
and the Belle~\cite{Hayasaka:2010np} collaborations, respectively: 
\begin{align}
 \mathrm{Br}(\mu^+ \to e^+ e^+ e^-) 
&< 1.0 \times 10^{-12}, 
\label{bound_sindrum}
\\
 \mathrm{Br}(\tau^-\to e^- e^+ e^-)
 &< 2.7 
\times 10^{-8}. 
 \label{belle_eee}
 \end{align}
The experimental limits (\ref{bound_sindrum}) and (\ref{belle_eee})
can be read as constraints on the LFV parameter $G_{ij}$ through
eqs.~(\ref{brmu3e}) and (\ref{brtau3l}).

In Fig.~\ref{fig:xsec_epem}, 
we show the cross sections for $e^+ e^- \to e^+ \ell^-$ (upper) 
and $e^- e^- \to e^- \ell^-$ (lower) from eqs.~(\ref{diffxepem}) and 
(\ref{diffxemem}) with the pseudo-rapidity $|\eta| \le 2.5$ as functions
of the LFV parameter $G_{12}$ (left), $G_{34}$ (center) and $G_{56}$
(right).
In each figure, the cross section is evaluated varying only $G_{ij}$
shown at the horizontal axis, and the other two parameters are set to be
zero.
The green, red and blue lines correspond to the CM energy
$\sqrt{s}=250~\mathrm{GeV}$, $500~\mathrm{GeV}$ and $1~\mathrm{TeV}$, 
respectively.
The dotted and dashed vertical lines denote 
constraints on $G_{ij}$ from SINDRUM for $\ell=\mu$ and Belle for
$\ell=\tau$, respectively. 

It can be seen from Fig.~\ref{fig:xsec_epem} that the LFV parameter
$G_{ij}$ for $\ell=\mu$ is strongly constrained from $\mu \to 3e$.
Taking account of the bounds on $G_{ij}$ for $\ell=\mu$, 
the cross sections are roughly smaller than $10^{-5}~\mathrm{fb}$ for
both $e^+ e^-\to e^+ \mu^-$ and $e^- e^-\to e^- \mu^-$, which are too
small to observe the LFV process at the ILC.
On the other hand, since current limits on $G_{ij}$ for $\ell=\tau$ are
much weaker than the $\mu$-channel, it could be expected that the ILC
experiment has a certain sensitivity for exploring the LFV processes
using the $\tau$-channel. 
We, therefore, focus on the $\ell=\tau$ case in the following
study\footnote{hereafter we suppress the index $\ell$ in $G^\ell_{ij}$.}.
It should be noted that 
the limit of the LFV decay $\tau \to 3\ell$
is expected to be improved significantly in the level of $O(10^{-10})$
at the super-KEKB~\cite{Bevan:2014iga}.  
\section{Constraints on the LFV parameters at the ILC}
Next we study constraints on the LFV parameter $G_{ij}$ expected at the
ILC experiments.
The SM background processes on the signal processes (\ref{signalepem}) and
(\ref{signalemem}) are 
\begin{align}
 e^+ e^- &\to e^+ \nu_e \tau^- \bar{\nu}_{\tau},
 \label{eq:bgepem}
 \\
 e^- e^- &\to e^- \nu_e \tau^- \bar{\nu}_{\tau}.
  \label{eq:bgemem}
\end{align}
To estimate the background events quantitatively, 
we generate these processes using　
{\sc MadGraph5\_aMC@NLO}~\cite{Alwall:2014hca} 
with the pseudo rapidity $|\eta| \le 2.5$.　
We summarize the cross section of the SM background processes 
 (\ref{eq:bgepem}) and (\ref{eq:bgemem}) 
 for 
 $\sqrt{s}=250~\mathrm{GeV},~500~\mathrm{GeV}$ and $1~\mathrm{TeV}$ 
in Table ~\ref{xsec_bg_unpol}. 
\begin{table}[h]
\begin{center}
 \begin{tabular}{@{\vrule width 1pt}l|c|c|c@{\ \vrule width 1pt}}
\hline \hline
  & $\sqrt{s}=250~\mathrm{GeV}$ & $500~\mathrm{GeV}$& $1~\mathrm{TeV}$\\[0.2mm]
\hline
  $\sigma(
 e^+ e^- \to e^+ \nu_e \tau^- \bar{\nu}_{\tau}
  )$~[fb]  & 203 & 113 &  85.5\\[0.2mm]
\hline
  $\sigma(
 e^- e^- \to e^- \nu_e \tau^- \bar{\nu}_{\tau}
  )$~[fb]  & 29.7 & 122 & 198\\
\hline \hline
 \end{tabular}
 \caption{
 Summary of 
 cross section of the background processes 
 (\ref{eq:bgepem}) and (\ref{eq:bgemem}) for 
 $\sqrt{s}=250~\mathrm{GeV},~500~\mathrm{GeV}$ and $1~\mathrm{TeV}$
 obtained by 
 {\sc MadGraph5\_aMC@NLO}~\cite{Alwall:2014hca}.  
}
 \label{xsec_bg_unpol}
\end{center}
\end{table}
Then the upper limits of the LFV parameters are examined using the 
$\chi^2$-function defined as 
\begin{align}
 \chi^2 \equiv \frac{(N_{\mathrm{S+B}}-N_\mathrm{B})^2}{N_\mathrm{B}
},   
\label{chisq}
\end{align} 
where subscripts S and B denote the signal and background respectively. 
The number of event $N$ is defined 
from the cross section $\sigma$ and the integrated luminosity
$\int dt \mathcal{L}$ 
as $N \equiv \sigma\cdot \int dt \mathcal{L}$.
In the following analysis, we use the set of $\sqrt{s}$ and the
luminosity at each phase of the ILC 
experiment~\cite{Behnke:2013xla} as summarized in Table~\ref{ilctable}. 
We set the 95\% CL limit on the LFV parameters at $\chi^2 = 3.84$. 
\begin{table}[h]
\begin{center}
 \begin{tabular}{@{\vrule width 1pt}l|c|c|c@{\ \vrule width 1pt}}
\hline \hline
  & (i) & (ii)& (iii)\\[0.2mm]
\hline
$\sqrt{s}~\mathrm{(GeV)}$ & 250& 500& 1000\\[0.2mm]
\hline
${\cal L}~(10^{34}~\mathrm{cm^{-2}~s^{-1}})$ & 
0.75& 3.6& 3.6\\[0.2mm]
\hline \hline
 \end{tabular}
\caption{summary of center-of-mass energy ($\sqrt{s}$) and 
luminosity planned at the ILC experiment~\cite{Behnke:2013xla}. 
}
 \label{ilctable}
\end{center}
\end{table}
\begin{figure}[t]
  \begin{center}
		 \includegraphics[clip, width=16cm]
		 {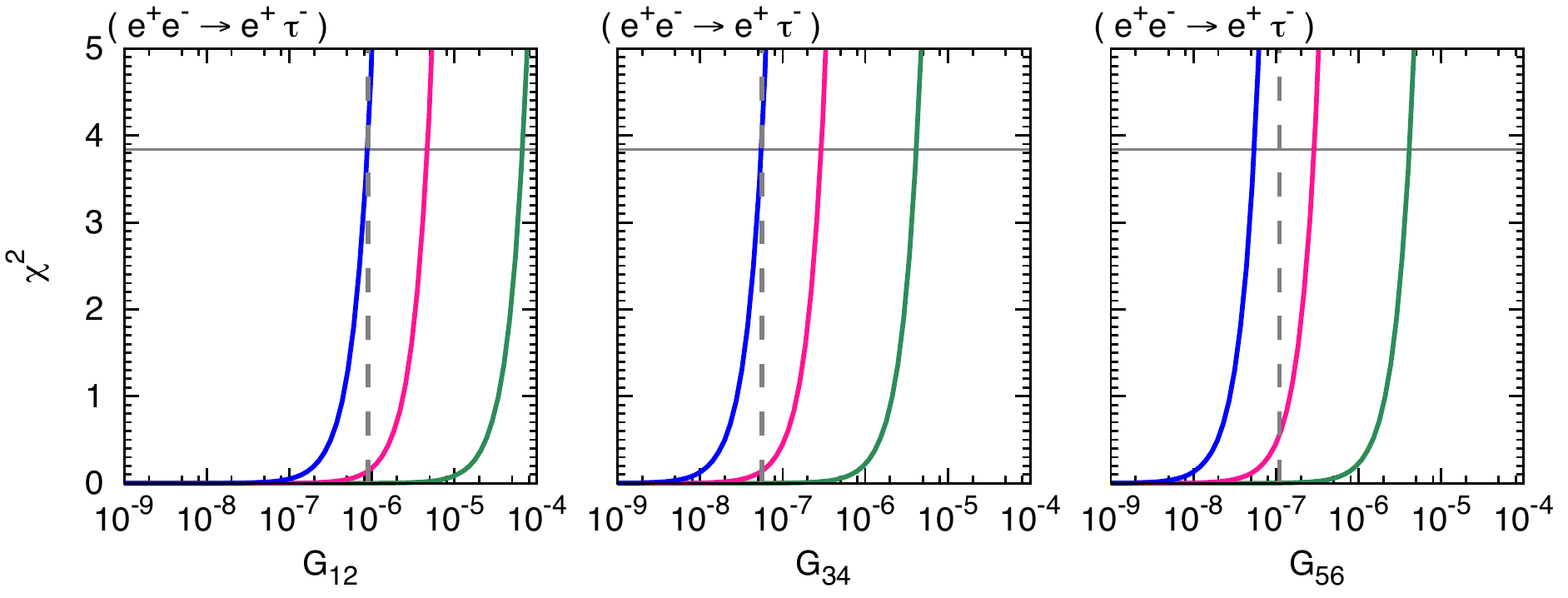}
		 \includegraphics[clip, width=16cm]
		 {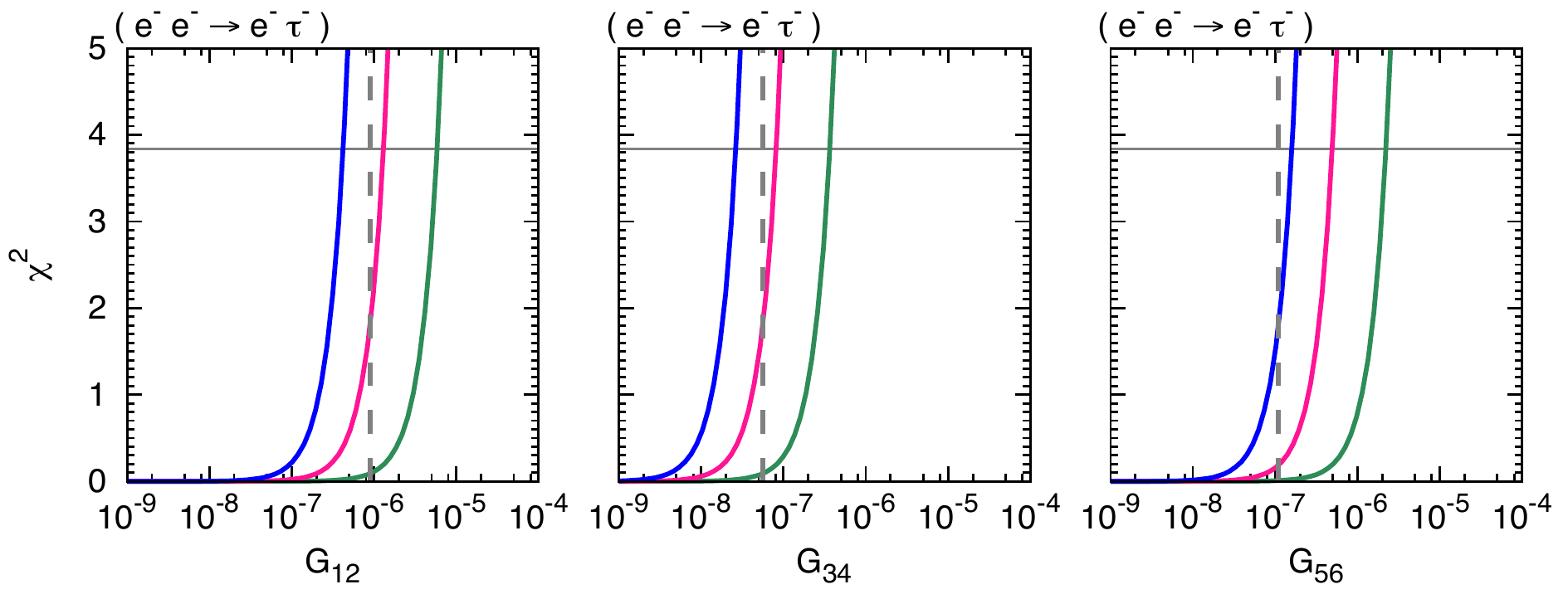}
    \caption{The $\chi^2$-parameters as functions of $G_{12},G_{34}$ and
   $G_{56}$ for $e^+ e^-$ (upper) and $e^- e^-$ (lower) collisions.
   The parameters (i), (ii) and (iii) in Table~\ref{ilctable} are used 
   to obtain green, red and blue curves, respectively. The vertical
   line denotes the upper bound on $G_{ij}$ from $\tau \to 3e$
   and the horizontal-dashed line represents $\chi^2=3.84$.
   }
    \label{fig:lena}
  \end{center}
\end{figure}

We show $\chi^2$ as a function of the LFV parameter for
$e^+ e^- \to e^+ \tau^-$ (upper) and $e^- e^- \to e^- \tau^-$ (lower) 
in Fig.~\ref{fig:lena}.
The green, red and blue curves are obtained using 
(i), (ii) and (iii) in Table~\ref{ilctable}, respectively.
The vertical dashed-line in each figure denotes 
the upper bound on $G_{ij}$ from $\mathrm{Br}(\tau \to 3e)$;
 \begin{align}
  G_{12} < 9.0\times 10^{-7},
  ~~
  G_{34} < 5.6\times 10^{-8},
  ~~
  G_{56} < 1.1\times 10^{-7}. 
\label{expboundtau}
 \end{align} 
The horizontal-dashed line represents $\chi^2=3.84$.
We find that, for both $\sqrt{s}=250~\mathrm{GeV}$ (i) and
$500~\mathrm{GeV}$ (ii), none of three LFV parameters are improved  
over the upper limits from $\mathrm{Br}(\tau \to
3e)$~(\ref{expboundtau}). 
For $\sqrt{s}=1~\mathrm{TeV}$, the upper limits at 95\% CL on $G_{56}$
in the $e^+ e^-$ collision and $G_{12}, G_{34}$ in the $e^- e^-$
collision are better 
than those from $\mathrm{Br}(\tau \to 3e)$~(\ref{expboundtau}), but the
improvement is marginal. 
\begin{figure}[t]
  \begin{center}
		 \includegraphics[clip, width=16cm]
		 {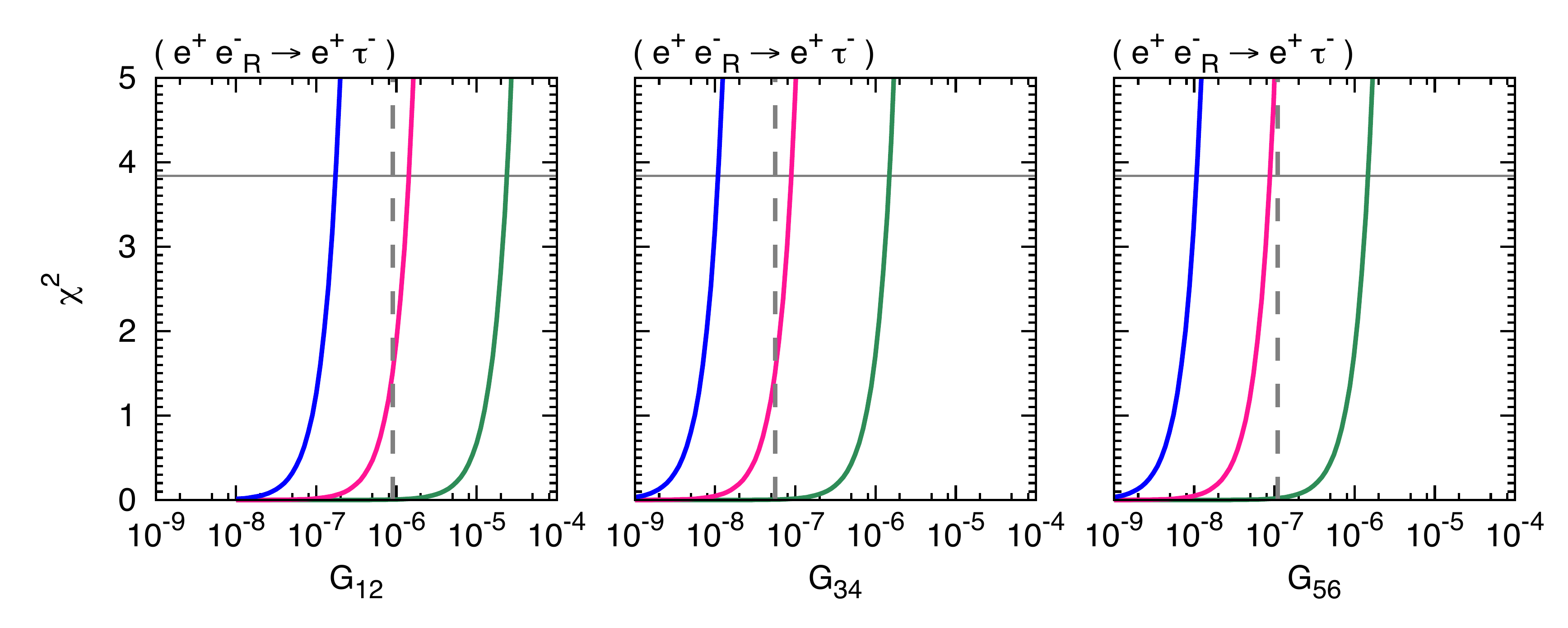}
		 \includegraphics[clip, width=16cm]
		 {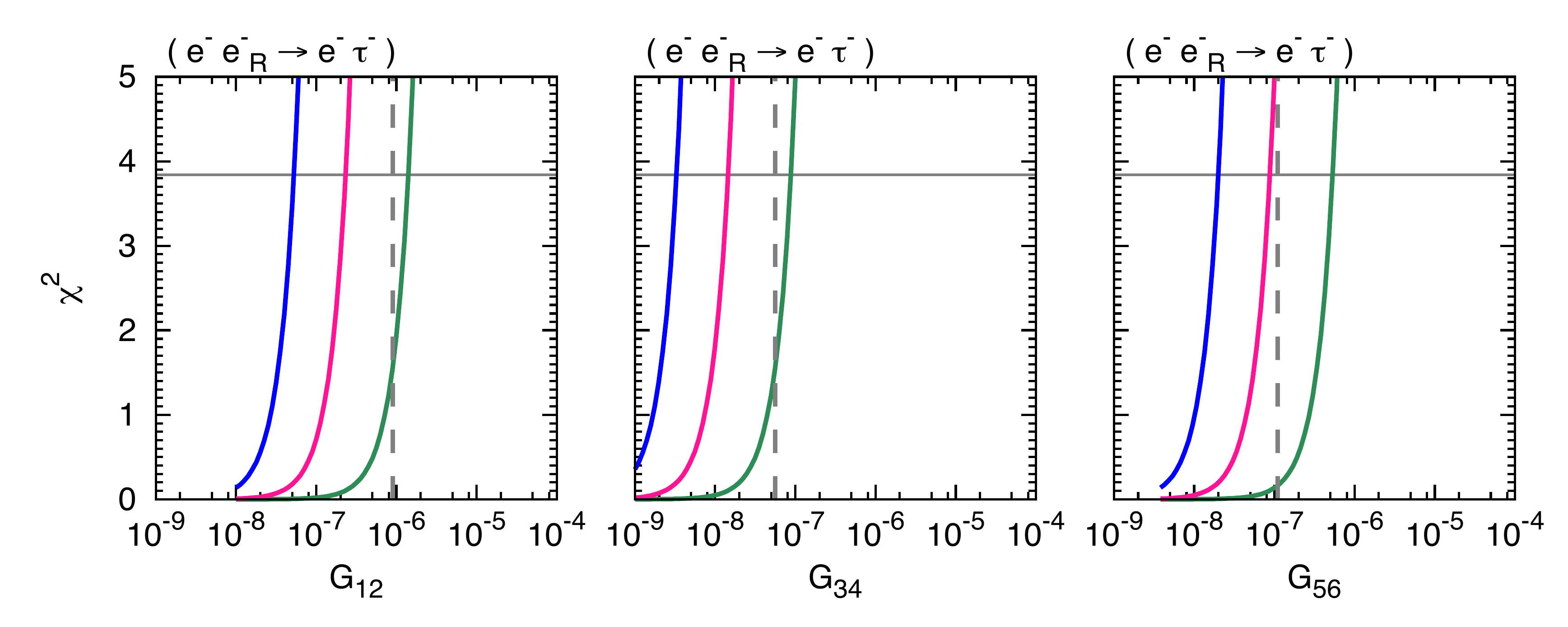}
    \caption{The $\chi^2$-parameters as functions of $G_{12},G_{34}$ and
   $G_{56}$ for $e^+ e_R^-$ (upper) and $e^- e^-_R$ (lower) collisions.}
    \label{fig:polepem}
  \end{center}
\end{figure}

Next we discuss a possibility to use the polarized electron beam.
In many of the Feynman diagrams for the background processes 
(\ref{eq:bgepem}) and (\ref{eq:bgemem}), the initial electron couples to 
the $W$-boson which appepars in the intermediate state. 
Contributions from such diagrams are suppressed when the electron beam 
is polarized to be right-handed, since only the left-handed electron
can couple to the $W$-boson. 
\begin{table}[h]
\begin{center}
 \begin{tabular}{@{\vrule width 1pt}l|c|c|c@{\ \vrule width 1pt}}
\hline \hline
  & $\sqrt{s}=250~\mathrm{GeV}$ & $500~\mathrm{GeV}$& $1~\mathrm{TeV}$\\[0.2mm]
\hline
  $\sigma(
 e^+ e^-_R \to e^+ \nu_e \tau^- \bar{\nu}_{\tau}
  )$~[fb]  & 3.45 & 0.827 & 0.256\\[0.2mm]
\hline
  $\sigma(
 e^- e^-_R \to e^- \nu_e \tau^- \bar{\nu}_{\tau}
  )$~[fb]  & 6.95 & 15.0 & 12.4\\
\hline \hline
 \end{tabular}
 \caption{
 Summary of 
 cross section of the background processes 
 (\ref{eq:bgepem}) and (\ref{eq:bgemem}) for 
 $\sqrt{s}=250~\mathrm{GeV},~500~\mathrm{GeV}$ and $1~\mathrm{TeV}$
  obtained by 
 {\sc MadGraph5\_aMC@NLO}~\cite{Alwall:2014hca}.  
}
 \label{xsec_bg_pol}
\end{center}
\end{table}
In Fig.~\ref{fig:polepem},
we show $\chi^2$ as a function of $G_{ij}$ for the $e^+ e^-_R$ 
(upper) and the $e^- e^-_R$ (lower) collisions.   
As we expected, the suppression of the SM background processes due to
the right-handed electron beam makes the upper limits on the LFV
parameter $G_{ij}$ better than the unpolarized case shown in
Fig.~\ref{fig:lena}.
For the $e^+ e^-_R$ collision, the 95\% CL upper limits on three LFV
parameters for $\sqrt{s}=1~\mathrm{TeV}$ are
\begin{align}
 G_{12}< 1.8\times 10^{-7},
 ~~
 G_{34}< 1.0\times 10^{-8}, 
 ~~
 G_{56}< 1.1\times 10^{-8},
 \label{limit1}
\end{align}
which are a few factors smaller than those from
$\mathrm{Br}(\tau\to 3e)$ at the Belle experiment~(\ref{expboundtau}).
On the other hand, the bounds on $G_{ij}$ for both
$\sqrt{s}=250~\mathrm{GeV}$ and $500~\mathrm{GeV}$ are still worse than
those in $\tau\to 3e$, except for $G_{56}$ for
$\sqrt{s}=500~\mathrm{GeV}$ which is comparable with
$\mathrm{Br}(\tau\to 3e)$.
In the $e^- e^-_R$ collision, the upper limits on $G_{ij}$ are 
improved further.
For $\sqrt{s}=500~\mathrm{GeV}$, the upper limits at 95\% CL are given
as
\begin{align}
 G_{12} < 2.2\times 10^{-7},
 ~~
 G_{34} < 1.3\times 10^{-8},
 ~~
 G_{56} < 8.5\times 10^{-8}, 
 \label{limit2}
\end{align}
and those for $\sqrt{s}=1~\mathrm{TeV}$ are given as 
\begin{align}
 G_{12} < 5.0\times 10^{-8},
 ~~
 G_{34} < 3.1\times 10^{-9},
 ~~
 G_{56} < 1.9\times 10^{-8}, 
 \label{limit3}
\end{align}
 which are better than the limits from $\mathrm{Br}(\tau \to
 3e)$~(\ref{expboundtau}).  

We have so far discussed the search limits on the LFV processes at the
ILC without taking account of the decay of the $\tau$-lepton in the 
final states.
The ILC is expected to achieve a good detector performance for
efficiency and purity of the main $\tau$-decay mode 
selections~\cite{Behnke:2013lya}, as shown in Table~\ref{taubr} together 
with the experimental data of branching ratio~\cite{Agashe:2014kda}.
The efficiency and purity in the table were calculated from 
$\tau^+ \tau^-$ production events, in which two $\tau$ candidates are
required to be almost back-to-back.
It is, therefore, a good approximation to assume 100\% efficiency of 
these decay modes in the following discussion. 
Using three decay modes in Table~\ref{taubr},  
we find that the upper bounds of the LFV 
parameter are twice (1/(0.178+0.174+0.108)=2.17) 
as large as (\ref{limit1}), (\ref{limit2}) and (\ref{limit3}).
For example, the upper limits at 95\% CL 
for $\sqrt{s}=1~\mathrm{TeV}$ in the $e^-
e^-_R$ collision are given as
\begin{align}
 G_{12} < 1.1\times 10^{-7},
 ~~
 G_{34} < 6.8\times 10^{-9},
 ~~
 G_{56} < 4.2\times 10^{-8}. 
 \label{limit4}
\end{align}
The improvement is nearly an order of magnitude in $G_{12}$ and 
$G_{34}$. 
\begin{table}[h]
\begin{center}
 \begin{tabular}{l|c|c|c}
  \hline \hline
  Mode  & Branching ratio~\cite{Agashe:2014kda} 	  
	  & Efficiency~\cite{Behnke:2013lya}
		  & Purity~\cite{Behnke:2013lya}
			  \\ \hline
  $e^- \bar{\nu}_e \nu_\tau$& 17.8\% & 98.9\% & 98.9\%
			  \\
  $\mu^- \bar{\nu}_\mu \nu_\tau$&17.4\%  & 98.8\% & 99.3\%
			  \\
  $\pi^- \nu_\tau$& 10.8\% & 96.0\% & 89.5\%
			  \\
\hline \hline
 \end{tabular}
 \caption{
Branching ratio, efficiency and purity of main $\tau$ decay modes. 
}
 \label{taubr}
\end{center}
\end{table}

\section{Summary}
In summary, we have investigated constraints on four-Fermi contact
interactions which leads to the LFV processes at the ILC experiments.
The cross sections for both $e^+ e^- \to e^+ \ell^-$ and $e^- e^- \to
e^- \ell^-$ are parametrized by three LFV parameters,
$G^\ell_{12},G^\ell_{34}$ and $G^\ell_{56}$, for each lepton flavor
$\ell$. 
Taking account of
constraints from measurements of
$\mathrm{Br}(\mu \to 3e)$ at the SINDRUM experiment
and 
$\mathrm{Br}(\tau \to 3e)$ at the Belle experiment,
we studied the upper limit on $G_{ij}$ expected at the ILC experiments. 
Although the ILC cannot give better bounds on the LFV parameters than
those at the SINDRUM experiment for $\ell=\mu$, we found that 
the measurements of cross sections of $e^+ e^- \to e^+ \tau^-$ and $e^-
e^- \to e^+ \tau^-$ at the ILC experiment could improve the upper limits
on the LFV parameters over those at the Belle experiment. 
In particular, the use of polarized electron beam increases the
sensitivity of measurements of the signal events by suppressing the SM
background significantly.
Owing to the expected high-efficiency of main $\tau$ decay modes at the
ILC,
we found that 
the 95\% CL upper limits on the LFV parameters 
at the ILC are in the level of $O(10^{-7}-10^{-9})$ 
for $\sqrt{s}=1~\mathrm{TeV}$,  
which are better than the previous experiment nearly an 
order of magnitude.

\section*{Acknowledgments}
The work of G.C.C is supported in part by Grants-in-Aid for Scientific 
Research 
from the Japan Society for the Promotion of Science (No.16K05314).  

 
%
\end{document}